# Competing Ground States in Triple-layered $Sr_4Ru_3O_{10}$: Verging on Itinerant Ferromagnetism with Critical Fluctuations

G. Cao[1], L. Balicas[2], W.H. Song[1*], Y.P. Sun[1*], Y. Xin[2], V.A. Bondarenko[1], J.W. Brill[1] and S. Parkin[3]

[1.] *Department of Physics and Astronomy, University of Kentucky, Lexington, KY 40506*

[2.] *National High Magnetic Field Laboratory, Tallahassee, FL 32310*

[3.] *Department of Chemistry, University of Kentucky, Lexington, KY 40506*

$Sr_4Ru_3O_{10}$ is characterized by a sharp metamagnetic transition and ferromagnetic behavior occurring within the basal plane and along the c-axis, respectively. Resistivity at magnetic field, B, exhibits low-frequency quantum oscillations when B∥c-axis and large magnetoresistivity accompanied by critical fluctuations driven by the metamagnetism when B⊥c-axis. The complex behavior evidenced in resistivity, magnetization and specific heat presented is not characteristic of any obvious ground states, and points to an exotic state that shows a delicate balance between fluctuations and order.

PACS numbers: 71.27+a, 75.47.-m

The layered ruthenates $(Ca,Sr)_{n+1}Ru_nO_{3n+1}$ with extended 4d-electron orbitals are characterized by the ground state instability. Physical properties of these correlated electron systems are critically linked to the lattice through the orbital degree of freedom, and thus exhibit an astonishing and distinctive dimensionality-dependence as the number of $RuO_6$ octahedral layers, n, changes [1-17]. The sensitivity of Fermi surface topography to the lattice is so strong that the ground state can be readily and drastically altered by the isoelectronic cation substitution [1-17].

$Sr_{n+1}Ru_nO_{3n+1}$, besides the celebrated p-wave superconductor $Sr_2RuO_4$ (n=1) [1], also includes the strongly enhanced paramagnet $Sr_3Ru_2O_7$ (n=2, $T_M$=18 K) [8], and the itinerant ferromagnet $SrRuO_3$ (n=∞, $T_C$=165 K) [11]. The bilayered $Sr_3Ru_2O_7$ demonstrates behavior consistent with proximity to a metamagnetic quantum critical point [9,10], and the infinite-layered $SrRuO_3$ is a robustly itinerant ferromagnet [8]. With n=3 intermediate between n=2 and ∞, the triple-layered $Sr_4Ru_3O_{10}$ (see inset Fig.1a) shows ferromagnetic behavior with $T_C$=105 K along the c-axis [17]. While the evolution of (ferro)magnetism in $Sr_{n+1}Ru_nO_{3n+1}$ apparently reflects variations in the Fermi surface topography consistent with the progression of n from 1 to ∞, intriguing transport and thermodynamic properties observed in this study suggest that $Sr_4Ru_3O_{10}$ is delicately poised between the itinerant metamagnetic state and the itinerant ferromagnetic state that characterize its closest neighbors $Sr_3Ru_2O_7$ and $SrRuO_3$, respectively. The anisotropic and complex magnetism is accompanied by critical fluctuations and equally anisotropic and complex transport behavior including large tunneling magnetoresistance and low frequency quantum oscillations that are uncommon in oxides and not discerned in $Sr_3Ru_2O_7$. In this paper, we report resistivity, magnetization and specific heat of single crystal $Sr_4Ru_3O_{10}$ as a function of temperature and magnetic field [18].

Displayed in Fig. 1 are the temperature dependences of the magnetization, M, resistivity, $\rho$, for the basal plane and c-axis, and specific heat, C, measured with ac-calorimetry [19,20] and normalized by a low-temperature value measured earlier [21]. In Fig.1a, M for the c-axis at B=0.01 T shows an abrupt Curie temperature, $T_C$, at 105 K, which is then followed by a sharp transition at $T_M$=50 K. The irreversibility of M becomes considerably large below $T_M$, consistent with ferromagnetic behavior. However, M for the basal plane exhibits only a weak cusp at $T_C$, but a pronounced peak at $T_M$, resembling antiferromagnetic-like behavior (see the inset), and showing no irreversibility. While the strong competition between ferromagnetic and antiferromagnetic coupling is always a major characteristic of the ruthenates, this behavior is not entirely expected in that the evolution of magnetism in $Sr_{n+1}Ru_nO_{3n+1}$ for n=1, 2, and $\infty$ would suggest an increasingly strong, but less anisotropic coupling favorable for ferromagnetism as n increases and as the orthorhombicity and the $RuO_6$ rotation are present [22,23].

The magnetic susceptibility for 150-350 K well obeys the Curie-Weiss law, yielding highly anisotropic Pauli susceptibility, $\chi_o$, of $4.1 \times 10^{-3}$ emu/mole and $1.4 \times 10^{-4}$ emu/mole for the c-axis and the basal plane, respectively. The largely enhanced $\chi_o$ for the c-axis suggests a large density of states near the Fermi surface, in accord with the Stoner criterion for ferromagnetism that occurs along the c-axis. On the other hand, $\chi_o$ for the basal plane is more than an order of magnitude smaller than that of the c-axis, implying a less energetically favorable condition for ferromagnetism so evident in Fig.1a. The surprisingly large anisotropic $\chi_o$ ($\chi_o^c/\chi_o^{ab}$=29) certainly points to the highly anisotropic density of states near the Fermi surface. Measurements of specific heat C(T) yields the electric specific heat $\gamma=109$ mJ/mol K$^2$. The Wilson ratio $R_w(=3\pi^2 k\chi_o/\mu_B^2\gamma)$ is found to be 2.4 and 0.08 for the c-axis and the basal plane, respectively, assuming $\gamma$ probes the average value of the renormalization effect over the entire Fermi surface. All these results reflect the high density of states for the c-axis that brings about the ferromagnetic behavior, and yet the much smaller density of states for the basal plane that is not large enough to cause ferromagnetic instability (see Table1 for comparisons). It is thus not entirely surprising that C(T) exhibits perfect $T^3$-dependence below 30 K without any visible deviation, e.g. $T^{3/2}$-dependence expected for (3D) ferromagnetic spin waves (see the inset of Fig.1b), although the low-temperature specific heat is currently being reexamined more closely [20].

Shown in Fig.1b is resistivity, $\rho$ vs. temperature for the c-axis and the basal plane. $\rho_c$ exhibits anomalies corresponding to $T_C$ and $T_M$ whereas $\rho_{ab}$ shows weaker ones at $T_C$ and $T_M$ (not obvious in the figures). The most prominent features are the unexpectedly large anisotropy and unusual temperature dependence of $\rho_c$. The ratio of $\rho_c/\rho_{ab}$ ranges from nearly 30 at 2 K to 10 at 350 K, a still quite two-dimensional characteristic despite of the triple layered structure. $\rho_c$ precipitously drops by nearly an order of magnitude from 50 K to 2 K. Such a drop in $\rho_c$ is somewhat similar to that of $Sr_2RuO_4$ and $Sr_3Ru_2O_7$ but more pronounced and less anticipated because the triple-layered structure should be more energetically favorable for inter-plane hopping. It is likely to be due to a drastic reduction of spin-scattering as the system becomes more spin-polarized below $T_M$ evidenced in Fig.1a. This is supported by the data shown in Fig.1c where the anomaly at $T_M$ rapidly decreases with increasing B applied parallel to the basal plane and eventually vanishes at B=4 T. The brief nonmetallic behavior (negative slope) seen between $T_M$(=50K) and 70 K may be associated with the elongated $RuO_6$ octahedra in the outer layers at low temperatures [17]. Our recent single crystal diffraction also shows that the c-axis is elongated by approximately 0.2% from 300 K to 90 K. Such a lengthened c-axis lifts the degeneracy of the $t_{2g}$ orbitals and narrows the $d_{xz}$ and $d_{yz}$ bands relative to the $d_{xy}$, weakening the inter-plane hopping. This point is in line with the high density of states indicated by $\chi_o^c$.

At low temperatures, Fermi liquid behavior survives as both $\rho_c$ and $\rho_{ab}$ obey $\rho = \rho_o + AT^2$ for the significant regime 2-15 K, with $\rho_{oc}$ = 1.30 x $10^{-3}$ $\Omega$ cm, $\rho_{oab}$ = 5.4 x $10^{-5}$ $\Omega$ cm (indicative of the high purity of the sample [18]), $A_c$ = 1.04 x $10^{-5}$ $\Omega$ cm/$K^2$ and $A_{ab}$ = 3.4 x $10^{-7}$ $\Omega$ cm/$K^2$. The c-axis values are very close to those of $Sr_3Ru_2O_7$ [8], but the basal plane values are about an order larger; i.e. while triple layer $Sr_4Ru_3O_{10}$ has a more isotropic resistivity than double layer $Sr_3Ru_2O_7$, it is not because of better transport along c but because of more basal plane scattering. It is also surprising that the anisotropies in A (~ 31) and $\rho_0$ (~ 24) are similar, since the latter depends only on the band mass and elastic scattering rate, while A depends on the interacting quasiparticle effective mass and inelastic rate, and, unlike $Sr_3Ru_2O_7$, the susceptibility is so anisotropic. In any case, all these parameters, tabulated in Table 1, clearly suggest that the Fermi surface remains very anisotropic in this triple layer material. The value of $A_c$ is much larger than that of ordinary Fermi liquids and, along with the c-axis Wilson ratio $R_W$ =2.4, suggests very strong correlation effects in this direction.

Consistent with the above results, both $\rho_c$ and $\rho_{ab}$ show neither $T^{3/2}$- nor $T^{5/3}$- dependence in the vicinity of the magnetic anomalies expected for a 3D antiferromagnet and ferromagnet, respectively [24,25]. Beyond the transition regions, $\rho_{ab}$ shows essentially linear temperature dependence for temperature ranges of 18-38 K, 50-100 K and 140-350 K, suggesting strong fluctuations. Furthermore, $\rho_{ab}$ shows significant negative magnetoresistivity, defined as $\Delta\rho/\rho(0T)$, that reaches as large as 28% in the vicinity of and below $T_M$ at B=5 T applied parallel to the basal plane (see Fig.1d). This is important because spin scattering is expected to be minimized at B=0 in itinerant ferromagnets where electrons find the path of least dissipation to cross the sample, and thus positive magnetoresistance is anticipated in the presence of B which forces electrons to take a different path. The observed negative magnetoresistance of $\rho_{ab}$ is a clear indication of large spin fluctuations existing in a non-ferromagnetic state. This point is consistent with the T-dependence of $\rho_{ab}$ at 5 T for 10-100 K shown in Fig.1d.

Table 1. Summary of some parameters for the c-axis and the basal plane

| Parameter | c-axis | Basal plane |
| --- | --- | --- |
| $\chi_o$ (memu/mole) | 4.1 | 0.13 |
| $\mu_{eff}$ ($\mu_B$/Ru) | 2.62 | 2.37 |
| Curie-Weiss temperature(K) | 122 | 118 |
| $\mu_s$ ($\mu_B$/Ru) (at 1.7K) | 1.13 | 0.9 |
| $\mu_{eff}/\mu_s$ | 2.32 | 2.63 |
| $B_c$ (T) (at 1.7 K) | 0.2 | 3.5 |
| $\rho_o$ ($\mu\Omega$ cm) | 1308 (0T) | 54 (0T) |
| A ($\mu\Omega$ cm/K$^2$) | 10.4 (0T) | 0.34 (0T) |
| $R_w$ | 2.4 | 0.08 |
| $A/\gamma^2$ ($\mu\Omega$-cm/(mJ/K$^2$-mol)$^2$) | 85.9×10$^{-5}$ | 2.8×10$^{-5}$ |

Indeed, as shown in Fig.2a and b where isothermal magnetization, M, for the basal plane and the c-axis for various temperatures is plotted, spins along the c-axis are readily polarized at only 0.2 T, yielding a saturation moment, $M_s$, of 1.13 $\mu_B$/Ru extrapolated to B=0 for T=1.7 K, slightly more than a half of 2 $\mu_B$/Ru expected for an S=1 system. This $M_s$ is comparable with that for $SrRuO_3$ whose easy axis lies within the basal plane [11], but the distinct field-dependence of M shown in Fig.2. is vastly different from that of $SrRuO_3$ [11] and $Sr_3Ru_2O_7$ [8]. While the irreversibility of M for the c-axis is typical of a ferromagnet, M for the basal plane shows a sharp metamagnetic transition, $B_c$. Clearly, $B_c$ decreases and broadens with increasing temperature, and vanishes at $T_M$=50 K (Fig.2b). The metamagnetism is observed in $Sr_3Ru_2O_7$ for both the basal plane and the c-axis with broader and higher $B_c$, which drives the ground state from a paramagnetic state at low fields to an induced ferromagnetic state at high fields [9,10]. Itinerant metamagnets are often characterized by a maximum in temperature dependence of paramagnetic susceptibility [26]. In light of the itinerant and the unusual magnetic character with ferromagnetism along the c-axis and antiferromagnetic-like behavior within the basal plane shown in Figs.1 and 2, it is intriguing as to whether this metamagnetic transition represents a transition between a paramagnetic and ferromagnetic state similar to that of itinerant metamgnets such as $Sr_3Ru_2O_7$, or a transition that characterizes a change from an antiferromagnetic state at low fields to a ferromagnetic state at higher fields, a process that manifests strong competing interactions between ferromagnetic and anitferromagnetic couplings, and is known to often occur in insulators [27]. It is clear, however, that this exotic behavior in $Sr_4Ru_3O_{10}$ is not characteristic of either a robust ferromagnet or an antiferromagnet, implying the close energies of the states.

This line of reasoning is reinforced by the fact that the c-axis transport properties appear to be dominated by the metamagnetic transition. Exhibited in Figure 2c are the magnetic field dependences of $A_c$ and $\rho_{0c}$, obtained by fitting the data below 15 K (see

Figure 1c) to $\rho_c = \rho_{oc} + A_c T^\alpha$, for $0 \leq B \leq 7T$. $\alpha = 2$, except between 2.7 and 2.9 T, for which the anomalous exponent $\alpha = 1.2$, as also found in $Ca_3Ru_2O_7$ [4] and in heavy Fermion compounds near quantum critical points [28]. The sharp peaking of both $A_c$ and $\rho_{oc}$ suggests large changes in carrier density and is strikingly similar to that in $Sr_3Ru_2O_7$ near 7.85 T below 0.35 K, the key evidence for its magnetic field-tuned quantum criticality [10].

The transport properties driven by the metamagnetism is also illustrated in Fig.3 where $\rho_c$ is plotted as a function of B rotating from the c-axis to the a-axis (no difference is discerned for B||a- and b-axis). $\Theta$ is defined as an angle between the c-axis and B. Noticeably, $\rho_c$ rises initially with increasing $\Theta$ by nearly 40% from 1.58 m$\Omega$ cm at $\Theta=0°$ (B||c or the easy axis) to 2.47 m$\Omega$ cm at $\Theta=42°$ at 32 T. This behavior suggests minimized scattering at $\Theta=0$ where spins are largely polarized along the c-axis. As $\Theta$ reaches $\Theta=43.4°$ $\rho_c$ starts to show negative megnetoresistivity characterized by a sudden drop at a critical field $B_c$. This $B_c$ systematically decreases with increasing $\Theta$, apparently corresponding to the metamagnetic transition that leads to the rapid spin polarization evidenced in M when B||ab. The abrupt metamagnetic transition yields a negative magnetoresistivity ratio of more than 60%. Such large inter-plane magnetoresistivity is believed to be due to a tunneling effect facilitated by a field-induced coherent motion of spin-polarized electrons between Ru-O planes, which is also seen in bilayered systems such as $Ca_3Ru_2O_7$ [3,4] and $La_{2-2x}Sr_{1+2x}Mn_2O_7$ [30]. Evidently, the metamagnetic transition $B_c$ reconstructs the Fermi surface and sharply divides the large negative magnetoresistivity and quantum oscillations, and the rapid disappearance of the SdH effect with changing $\Theta$ reflects the quite 2D nature of the Fermi surface.

The SdH effect expectedly becomes stronger as temperature decreases as shown in Fig.4a where $\rho_c$ vs. B with $\Theta=28º$ is plotted. Shown in Fig. 4b is the amplitude of the SdH signal [31] as a function of $B^{-1}$. As can be seen the detected frequency is low, and only a few oscillations are observed in the entire field ranging up to 32 T. The data analyses [31] reveal a frequency $F = 123 \pm 5$ T, which, based on the crystallographic data of $Sr_4Ru_3O_{10}$ [17] and the Onsager relation $F_0 = A(h/4\pi^2 e)$ ($e$ is the electron charge), corresponds to an area of only 0.9% of the first Brillouin zone. The inset shows the amplitude of the SdH signal measured at the highest fields as a function of temperature. The solid line is a fit to the Lifshitz-Kosevich formulae, $x/\sinh x$, where $x = 14.69\, \mu_c T/B$ [32]. The nearly perfect fit yields a cyclotron effective mass $\mu_c = 0.9 \pm 0.15$. The cyclotronic effective mass smaller than expected is attributed to the fact that it has been difficult to resolve the heavier effective masses in quantum oscillation experiments [4,33].

What makes this triple layered ruthenate intriguing is that it shows a delicate balance between fluctuations and order. This study reveals sound evidence for critical fluctuations driven by the metamagnetic transition in one direction and yet ferromagnetic behavior in the other. The strongly anisotropic transport properties seem to be critically linked to the metamagnetic transition that leads to the Fermi surface reconstruction evidenced in the quantum oscillations and the large magnetorsistance. All results presented suggest novel behavior that is characterized by non-characteristic ground state with proximity to the itinerant ferromagnetic instability and the possible quantum critical point. $Sr_4Ru_3O_{10}$ with readily tunable parameters opens a new avenue to study itinerant ferromagnetism and quantum criticality so imperative to highly correlated electrons.

Acknowledgements: We are grateful to Prof. Ganpathy Murthy for very helpful discussions. This work was supported in part by NSF grant DMR-0240813 (GC), a grant from University of Kentucky (GC), NSF grant DMR-0100572 (JWB) and NHMFL under cooperative agreement DMR-0084173 (YX).

* Permanent address: Institute of Solid State Physics, Chinese Academy of Sciences, Hefei 230031, Anhui, P.R. China.

Captions:

Fig.1. (a) Magnetization as a function of temperature for the basal plane and the c-axis at B=0.01 T. Inset: Enlarged M (T) for the basal plane below 150 K without axes to clarify the antiferromagnetic-like behavior; the TEM image and diffraction pattern taken along [110] direction shown illustrate the triple layered feature; (b) Basal-plane $\rho_{ab}$ and inter-plane resistivity $\rho_c$ as a function of temperature Inset: specific heat normalized by T, C/T, vs. $T^2$ for T≤31 K; (c) $\rho_c$ vs. T at a few representative magnetic fields, B, parallel to the basal plane; (d) $\rho_{ab}$ vs. T at B=0 and 5 T parallel to the basal plane.

Fig.2. (a) Isothermal magnetization as a function of B at T=1.7 K for the basal plane and the c-axis; (b) Basal plane isothermal magnetization at T=1.7, 15, 30, 50, 90, and 120 K. Note that metamagnetic transition decreases with increasing T and vanishes at T>50 K; (c) Magnetic field dependence of the coefficient A and residual $\rho_o$ obtained from the data shown in Fig.1c.

Fig.3. $\rho_c$ at T=0.6 K as a function of B rotating from the c-axis to the basal plane.

Fig.4. (a) $\rho_c$ as a function of B at $\Theta=28°$ for a few representative temperatures; (b) Shubnikov-de Hass effect as a function of $B^{-1}$ for various temperatures specified. Inset: Temperature dependence of the SdH magnitude normalized by T yields the cyclotron effective mass, $\mu_c$.

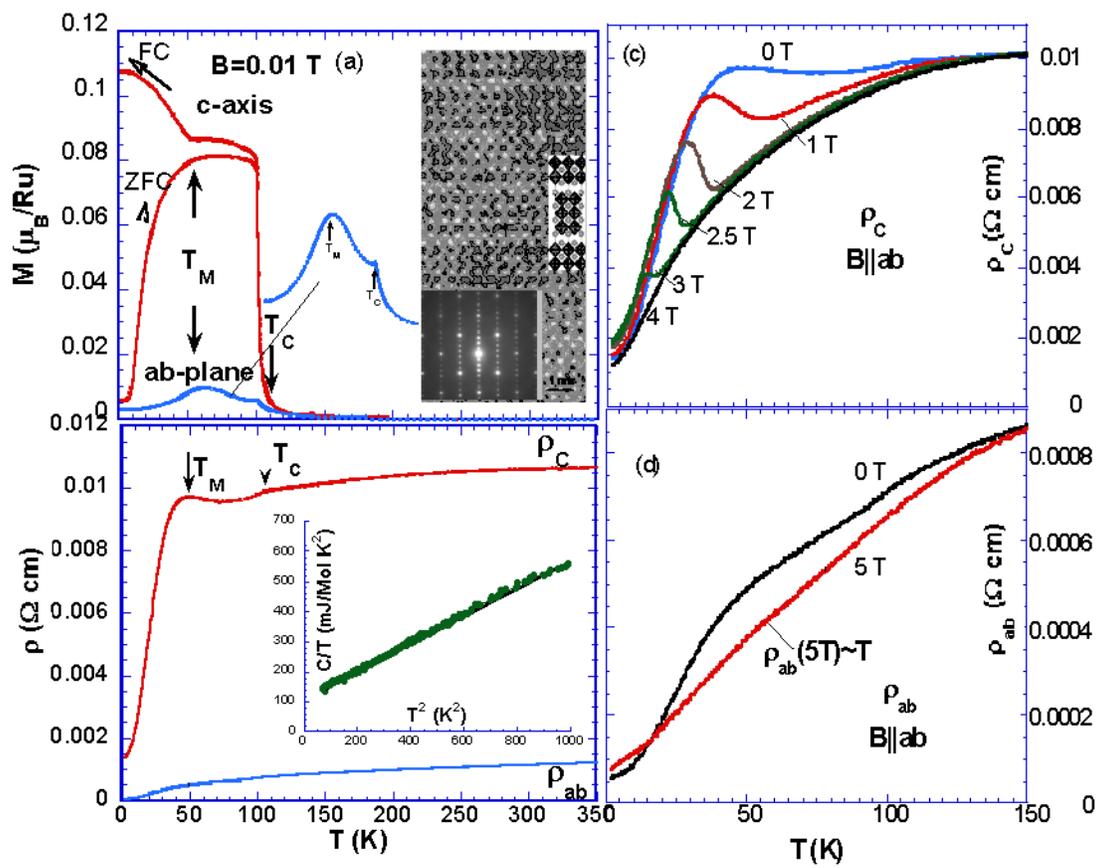

Fig.1, cao

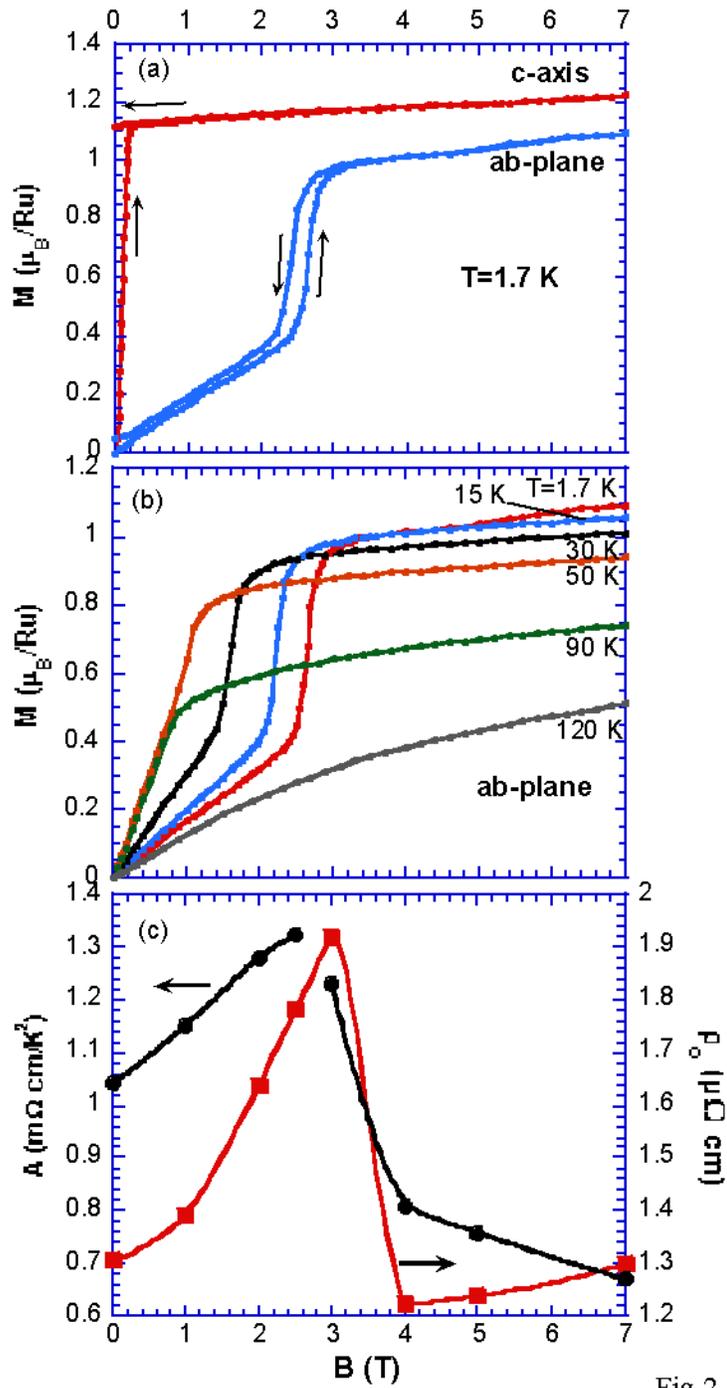

Fig.2, Cao

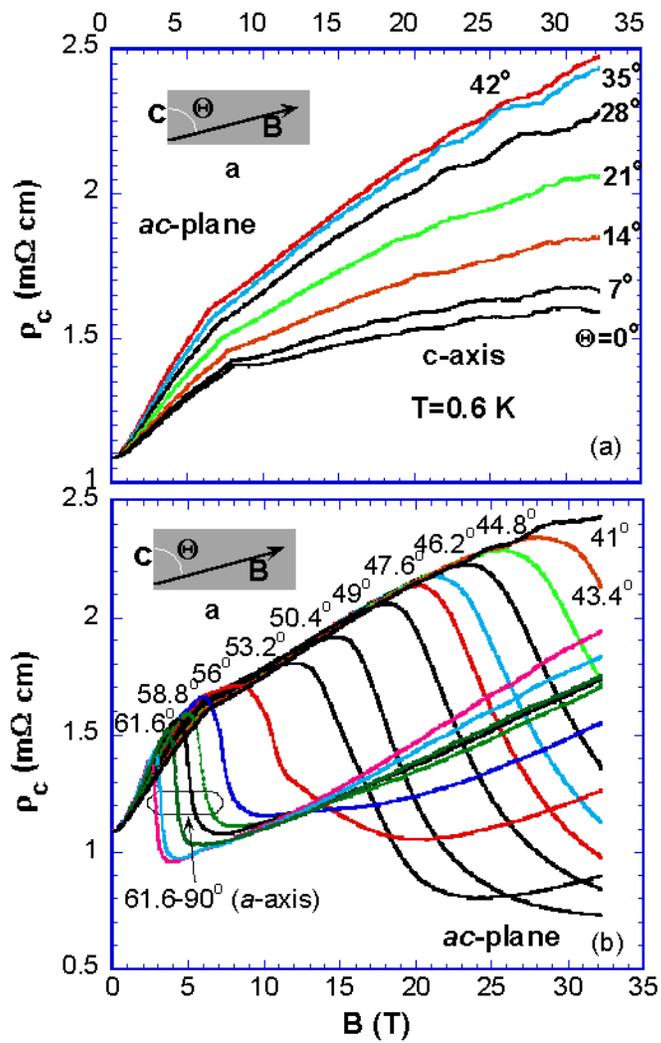

Fig.3, Cao

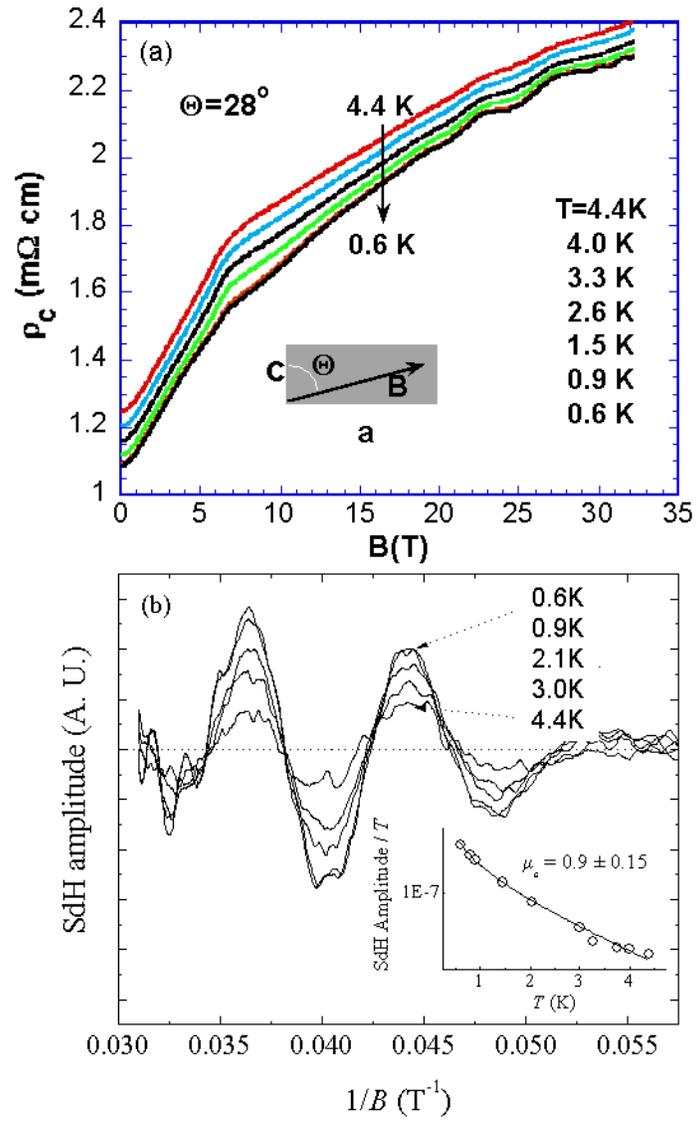

Fig. 4, Cao